\documentclass[letter]{aa}
\usepackage{graphicx}
\usepackage{txfonts}
\usepackage{natbib}
\usepackage{gensymb}
\usepackage[colorlinks=true,linkcolor=blue,citecolor=blue,urlcolor=blue]{hyperref}

\begin{document}

\newcommand{\reb}{{\sc \tt REBOUND}}
\newcommand{\whf}{{\sc \tt WHFast}}
\newcommand{\emcee}{{\sc \tt emcee}}

\title{The HARPS search for southern extra-solar planets}

\subtitle{XLII. A system of Earth-mass planets around the nearby M dwarf YZ~Cet
\thanks{Based on observations made with the HARPS instrument 
  on the ESO 3.6 m telescope under the program 
  IDs 180.C-0886(A), 183.C-0437(A), and 191.C-0873(A) at Cerro La Silla (Chile).}
\textsuperscript{,}\thanks{Radial velocity data (Table~\ref{tab:rv}) is only 
available at the CDS via anonymous ftp to 
cdsarc.u-strasbg.fr (130.79.128.5)\newline
or via\newline
http://cdsweb.u-strasbg.fr/cgi-bin/qcat?J/A+A/
}}

\authorrunning{Astudillo-Defru et al.}
\titlerunning{Multi-planetary system of Earth-mass planets within 3.6 pc}

\author{N. Astudillo-Defru\inst{\ref{geneva}}, 
R. F. D\'iaz\inst{\ref{uba},\ref{iafe}},
X. Bonfils\inst{\ref{grenoble}},
J.M. Almenara\inst{\ref{geneva}}, 
J.-B. Delisle\inst{\ref{geneva}},
F. Bouchy\inst{\ref{geneva}}, 
X. Delfosse\inst{\ref{grenoble}},
T. Forveille\inst{\ref{grenoble}},
C. Lovis\inst{\ref{geneva}}, 
M. Mayor\inst{\ref{geneva}}, 
F. Murgas\inst{\ref{iac}},
F. Pepe\inst{\ref{geneva}}, 
N. C. Santos\inst{\ref{caup}, \ref{porto}}, 
D. S\'egransan\inst{\ref{geneva}},
S. Udry\inst{\ref{geneva}}, 
A. W\"unsche\inst{\ref{grenoble}}}

\institute{Observatoire de Gen\`eve, Universit\'e de Gen\`eve, 
  51 ch. des Maillettes, 1290 Sauverny, Switzerland \label{geneva}\\ \email{nicola.astudillo@unige.ch}
  \and Universidad de Buenos Aires, Facultad de Ciencias Exactas y Naturales. Buenos Aires, Argentina.\label{uba}
  \and CONICET - Universidad de Buenos Aires. Instituto de Astronom\'ia y F\'isica del Espacio (IAFE). Buenos Aires, Argentina.\label{iafe}
    \and Univ. Grenoble Alpes, CNRS, IPAG, F-38000 Grenoble, France \label{grenoble} 
    \and Instituto de Astrof\'isica e Ci\^encias do Espa\c{c}o,
    Universidade do Porto, CAUP, Rua das Estrelas, PT4150-762 Porto,
    Portugal \label{caup}
    \and Departamento de F\'isica e Astronomia, Faculdade de Ci\^encias, 
    Universidade do Porto, Portugal \label{porto}
    \and Instituto de Astrof\'isica de Canarias, V\'ia L\'actea s/n, E-38205 La Laguna, Tenerife, Spain \label{iac}
    }

\date{}

\abstract{Exoplanet surveys have shown that systems with multiple low-mass
planets on compact orbits are common. Except for a few cases, however, the
masses of these planets are generally unknown. At the very end of the main
sequence, host stars have the lowest mass and hence offer the largest
reflect motion for a given planet. In this context, we monitored the
low-mass (0.13M$_\odot$) M dwarf YZ~Cet (GJ~54.1, HIP~5643) intensively
and obtained radial velocities and stellar-activity indicators
derived from spectroscopy and photometry, respectively. We find strong evidence
that it is orbited by at least three planets in compact orbits
(P$_{Orb}$=1.97, 3.06, 4.66 days), with the inner two near a 2:3
mean-motion resonance. The minimum masses are comparable to the
mass of Earth 
(M $\sin$i=0.75$\pm$0.13, 0.98$\pm$0.14, and 1.14$\pm$0.17 M$_\oplus$), and they are
also the lowest masses measured by radial velocity so far. We note the
possibility for a fourth planet with an even lower mass of 
M $\sin$i=0.472$\pm$0.096 M$_\oplus$ at P$_{Orb}$=1.04 days. An n-body
dynamical model is used to place further constraints on the system
parameters. At 3.6 parsecs, YZ~Cet is the nearest multi-planet system
detected to date.}

\maketitle

\section{Introduction}
\label{sec:introduction}

In the last 20 years, we have learned much about the vast diversity in the
architecture of planetary systems
\citep[e.g.,][]{Mayor2014,2014ApJ...790..146F}. 
For stars of different spectral types we have discovered planets
with a a large
variety in mass and size, constraining our understanding of
planetary formation processes.

M~dwarfs make up about 70\% of the stellar population of the Galaxy and
constitute the lower tail of the main sequence in the
Hertzsprung–Russell diagram. The detection of planets orbiting these
stars is therefore important to constrain the population of planets
and the formation processes in the less-massive protoplanetary disks 
\citep{Gaidos2017}.
Additionally, low-mass and small-sized M~dwarfs have advantages when
searching for the lowest-mass and smallest-sized planets: the radial
velocity (RV) amplitude and transit depth scales with M$_\star^{-2/3}$
and R$_\star^{-2}$, respectively.

It is known that M~dwarfs have a high occurrence of super-Earths, 
as shown from high-precision radial velocity   
\citep[f$\sim$0.88 for P<100 days
and M$\sin i$=1-10M$_\oplus$,][]{Bonfils2013} and transit surveys \citep[f$\sim$2.5 for P<200 days and
R=1-4R$_\oplus$,]
{Dressing2015}. Thanks to the most precise velocimeters combined
with the high amount of collected data, we are now starting to census planets
with minimum masses closer to the Earth mass such as GJ~273~c
\citep{Astudillo2017b}, Prox~Cen~b \citep{2016Natur.536..437A}, or 
GJ~628~b \citep{2016ApJ...817L..20W}.

The detection of multi-planetary systems sheds light on the theories
of model formation and evolution. Statistical studies have shown that
planet pairs near or in mean resonances are slightly favored
\citep[e.g.,][]{2014ApJ...790..146F}. In the understanding of this
scenario, a low uncertainty in the measured mass is important, but
the majority of the systems has been detected through transits, where
uncertainties are generally too large for a reliable physical
characterization of the planets 
\citep[e.g., Kepler-42 and Trappist-1,][respectively]{2012ApJ...747..144M,2017Natur.542..456G}.

In this paper we present the new nearby system (3.6 pc) of a
very-low-mass star that is orbited by at least three Earth-mass planets on
compact orbits, with the possibility of a fourth sub-Earth-mass
companion.

\section{Stellar properties and observations}
\label{sec:observations}

\subsection{Stellar properties}
\label{subsec:prop}

The properties of the mid-M~dwarf YZ~Cet are given in 
Table~\ref{tab:stellarprop}. The table lists the visual and near -infrared apparent magnitudes, the parallax and proper motions, basic
physical parameters, the activity level, and the age. These values come from 
the literature as detailed in the table caption.

\begin{table}[h]
  \caption{Stellar properties.}
  \label{tab:stellarprop}
  \centering
  \small
  \begin{tabular}{p{0.2\linewidth}l c c}  
    \hline\noalign{\smallskip}
    R.A.$^{\mathrm{1}}$  & [J2000] & $01^h 12^m 30.5^s$  \\
    Decl.$^{\mathrm{1}}$ & [J2000] & $-16\degree 59^\prime 56^{\prime\prime}$ \\
    Spectral type$^{\mathrm{1}}$    &  &  M4.5 \\
    V$^{\mathrm{1}}$ & &12.074     \\
    J$^{\mathrm{2}}$ & &7.258$\pm$0.020     \\          
    K$^{\mathrm{2}}$ & &6.420$\pm$0.017     \\
    $\pi$$^{\mathrm{3}}$ &  [mas]   &  271.01$\pm$8.36   \\
    $\mu_\alpha$$^{\mathrm{3}}$ & [mas/yr] & 1208.53$\pm$5.57 \\
    $\mu_\delta$$^{\mathrm{3}}$ & [mas/yr] &  640.73$\pm$3.71 \\
    $dv_r/dt$ &[m/s/yr] & 0.159$\pm$0.006  \\
    M$^{\mathrm{1}}$ & [M$_\odot$] & 0.130$\pm$0.013 \\
    R$^{\mathrm{1}}$ & [R$_\odot$] & 0.168$\pm$0.009 \\
    T$_{\rm eff}^{\mathrm{1}}$ & [K] & 3056$\pm$60 \\
    $[\mathrm{Fe/H}]^{\mathrm{1}}$ &  & -0.26$\pm$0.08  \\
    ${log(R^\prime_{HK})}^{\mathrm{4}}$ &  & -4.71$\pm$0.21 \\
    Age$^{\mathrm{5}}$ &  [Gyr]   &  5.0$\pm$1.0   \\
    \noalign{\smallskip}\hline
  \end{tabular}

  \begin{list}{}{}
  \item[$^1$] \citet{2015ApJ...804...64M,2016ApJ...819...87M}; 
  $^2$ \citet{2003yCat.2246....0C}; 
  $^3$ \citet{2007A&A...474..653V}; 
  $^4$ \citet{Astudillo2017a};
  $^5$ \citet[][assuming P$_{\mathrm{Rot}}$ = 83$\pm$15 d]{Engle2011}
  \end{list}
\end{table}

\subsection{HARPS radial velocities}
\label{subsec:rv}

We observed YZ~Cet from December 2003 to October 2016 with the 
HARPS spectrograph installed at the ESO 3.6m telescope in Chile 
\citep{2003Msngr.114...20M}. A total of 211 spectra were acquired 
with an exposure time of 900 s and fast read-out mode, corresponding
to a typical signal-to-noise ratio per spectra of 19 at 550 nm. Data 
were recorded without simultaneous wavelength reference to keep
stellar low-flux zones free of lamp contamination. The HARPS vacuum
vessel was opened in May 2015 to upgrade the fiber link
\citep[][a superscript plus indicates throughout data
acquired after the upgrade]{2015Msngr.162....9C}. This modifies the
line-spread function and may affect the spectral analysis. Consequently,
for some of our analyses, we independently analyze the pre- and post-fiber
upgrade data sets.

\subsection{ASAS photometry}
\label{subsec:phot}

This work uses data from the All Sky Automated Survey (ASAS) \citep{1997AcA....47..467P}. 
The ASAS project constantly monitors a large area of the sky 
($9\degree \times 9\degree$) to search for photometric variability in 
V and I band. The ASAS photometric accuracy is 0.01--0.02
mag for a 13$^{th}$ magnitude star, which is enough to detect spots
crossing the stellar disk due to rotation. 

The ASAS data of YZ~Cet consist of 459 photometric measurements
acquired from November 2000 to November 2009. Unfortunately, ASAS and
HARPS data do not overlap in time, which is desired to
understand RV variability that is not produced by planets. We used the
two-pixel wide aperture (MAG\_0), as suggested by ASAS for a 12.07
Vmag star.

\section{Analysis}
\label{sec:analysis}

\subsection{Periodogram+Keplerian}
\label{subsec:kep}

\begin{figure}[t]
\centering
\includegraphics[scale=0.45]{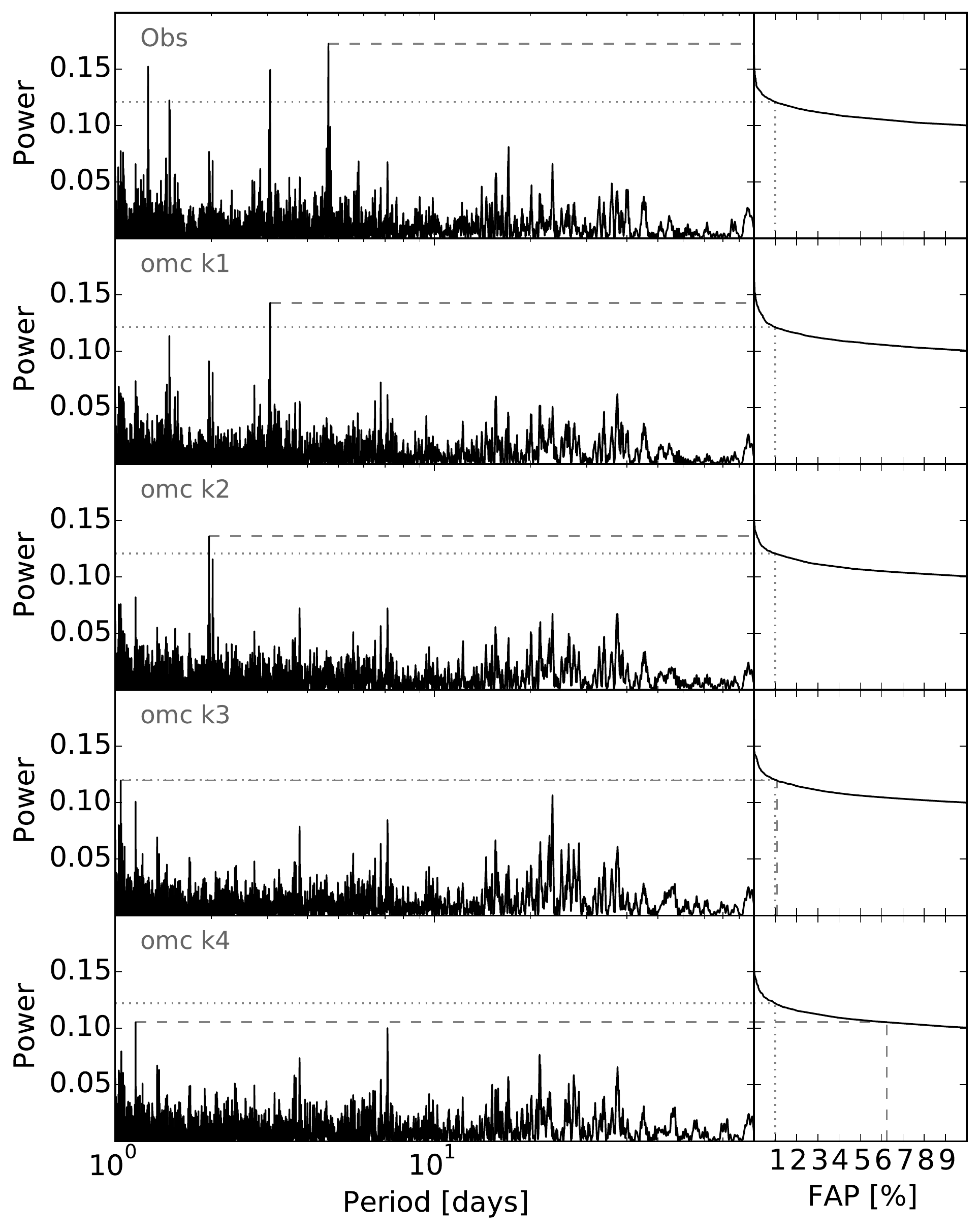}
\includegraphics[scale=0.45]{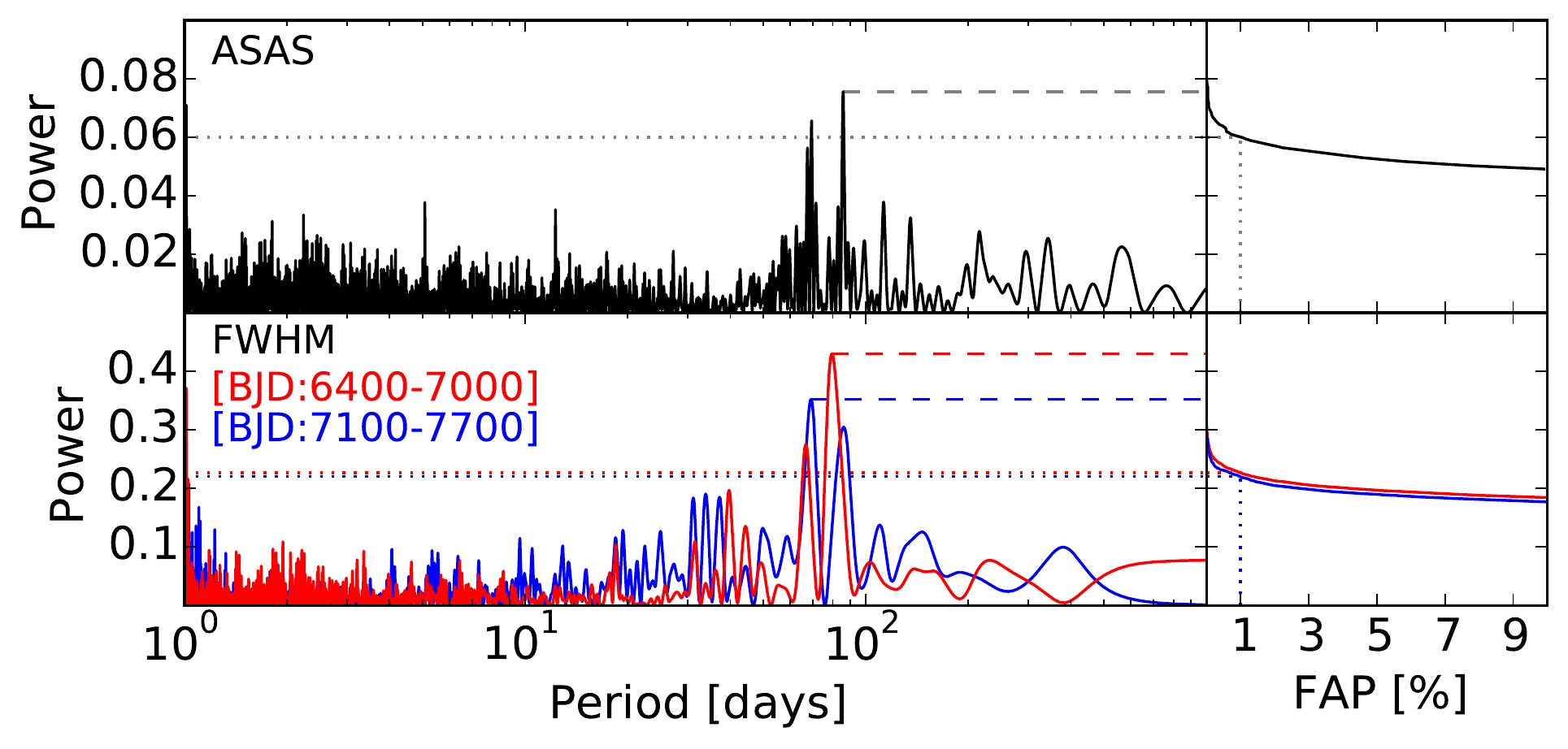}
\caption{Periodogram analysis of the RV (top panel) and the ASAS
photometry and FWHM$_{CCF}$ (bottom panel). Dashed lines highlight
the strongest peaks, dotted lines represent the 1\% FAP threshold. 
Three RV signals have an FAP lower than 1\%, while a fourth is just above this 
level of confidence. The final RV residual (omc k4) does not show any 
significant periodicity. The photometry and FWHM$_{CCF}$ show a
$\sim$83-day periodicity linked to the stellar rotation. Pre- (red) 
and post- (blue) fiber upgrade data are analyzed separately. }
\label{fig:rv_period}
\end{figure}

We computed the generalized Lomb-Scargle periodogram
\citep{Zechmeister2009} of the RVs and residues by progressively
subtracting a Keplerian fit accounting for the strongest peak in the
periodogram (Fig.~\ref{fig:rv_period}). 
We adjusted up to four Keplerian curves plus a velocity offset 
between the data obtained before and after the HARPS fibre upgrade 
\citep[BJD=2457172,][]{2015Msngr.162....9C}.
The top panel in Figure~\ref{fig:rv_period} shows three clear periodic
RV signals at 4.7, 3.1, and 2.0 days, with a false-alarm probability
(FAP) <1/10,000. A fourth less clear signal is seen with a one-day 
periodicity and a 1.1\% FAP. The Bayesian information criterion
(BIC) for a constant model is  857.83, and for a model of one, two, three, 
and four Keplerian, the BIC is 463.57, 420.21, 392.62, and
376.10, respectively. On the one hand, the BIC favors the model consisting
of four Keplerian curves, but on the other hand, the FAP of the fourth
signal is above the 1\% FAP threshold. We prefer to stay on the
conservative side and consider the fourth signal as tentative. A more
rigorous Bayesian model comparison and/or additional data are needed to
elucidate the reliability of the one-day RV signature.

\subsection{Stellar activity}
\label{subsec:act}

YZ~Cet is an active star with $log(R^\prime_{HK})$=-4.705$\pm$0.208. 
Such a $R^\prime_{HK}$ level translates into a rotation period of 
27$\pm$9 days \citep{Astudillo2017a}. However, there are clues that
flaring stars like Proxima Centauri do not follow the
$log(R^\prime_{HK})-P_{Rot}$ relationship. This is the case for 
YZ~Cet, where several flares are identified in spectra through 
Ca \ion{II} {HK}, Na D, He \ion{I} {D3}, and 
H$\alpha \beta\gamma \delta$. These chromospheric activity indicators
do not show evidence of stellar rotation.

Contrary to the chromospheric activity indicators, the V-band
photometry and the FWHM show evidence of the rotation period of
YZ~Cet ($\sim$83 days, Fig.~\ref{fig:rv_period}). Conversely,
none of the periodicities detected in the RV time series
appears in the photometry or FWHM data. 

\subsection{Three Keplerian+Gaussian process}
\label{subsec:gp}

\newcommand{\struct}{\epsilon}
\newcommand{\per}{\mathcal{P}}

As the star rotates, quasi-periodic signals may be introduced in RV 
as a result of evolving active regions. These signals are often very complex,
and a physical model is therefore usually hard to produce or has the
strong disadvantage of being computationally expensive. Gaussian
process regression (GP) has become a widely used non-parametric
method to model this variability without the necessity of specifying a model
\citep{RasmussenWilliams2006,Haywood2014, Rajpaul2015, Faria2016},
although it may be prone to inducing false positives if care is not taken
\citep{Dumusque2017}. 

A GP with a quasi-periodic covariance function was added to the
three-Keplerian model. The GP hyperparameters were sampled together with
the model parameters. We set a normal prior on the recurrence timescale,
$\per$ (see Appendix~\ref{sec:GP}), centered at 83 days and with a width of
15 days, in agreement with the measured stellar rotation rate (see
Sect.~\ref{subsec:act}). A detailed description of the model and the
choice of priors is provided in Appendix~\ref{sec:GP}.

The 23-dimensional parameter space of our model was explored with 
the Markov chain Monte Carlo (MCMC) algorithm described in \citet{goodmanweare2010} and
implemented by \citet{emcee}, where 100 walkers were used. The
algorithm was started at a random point in the prior and was allowed
to evolve until no further variation in mean
posterior density across the walkers was seen. Then, the algorithm
was run for a further 100,000 steps used for the inference process.
We retained only those walkers that explored the main period peak for
each planet (Fig~\ref{fig:rv_period}), whose samples where used to initialize a new 70,000-step MCMC run using 300 walkers. The final inference
was made on $2.1\times10^7$ posterior samples, out of which over 21,000 are independent (see Appendix~\ref{sect:MCMC} for details). Results are reported in
Table~\ref{tab:OrbitalParam}.

\section{Results}
\label{sec:result}

\subsection{Compact system near the 3:2 resonances}
\label{subsec:planets}

\begin{figure}
\centering
\includegraphics[scale=0.92]{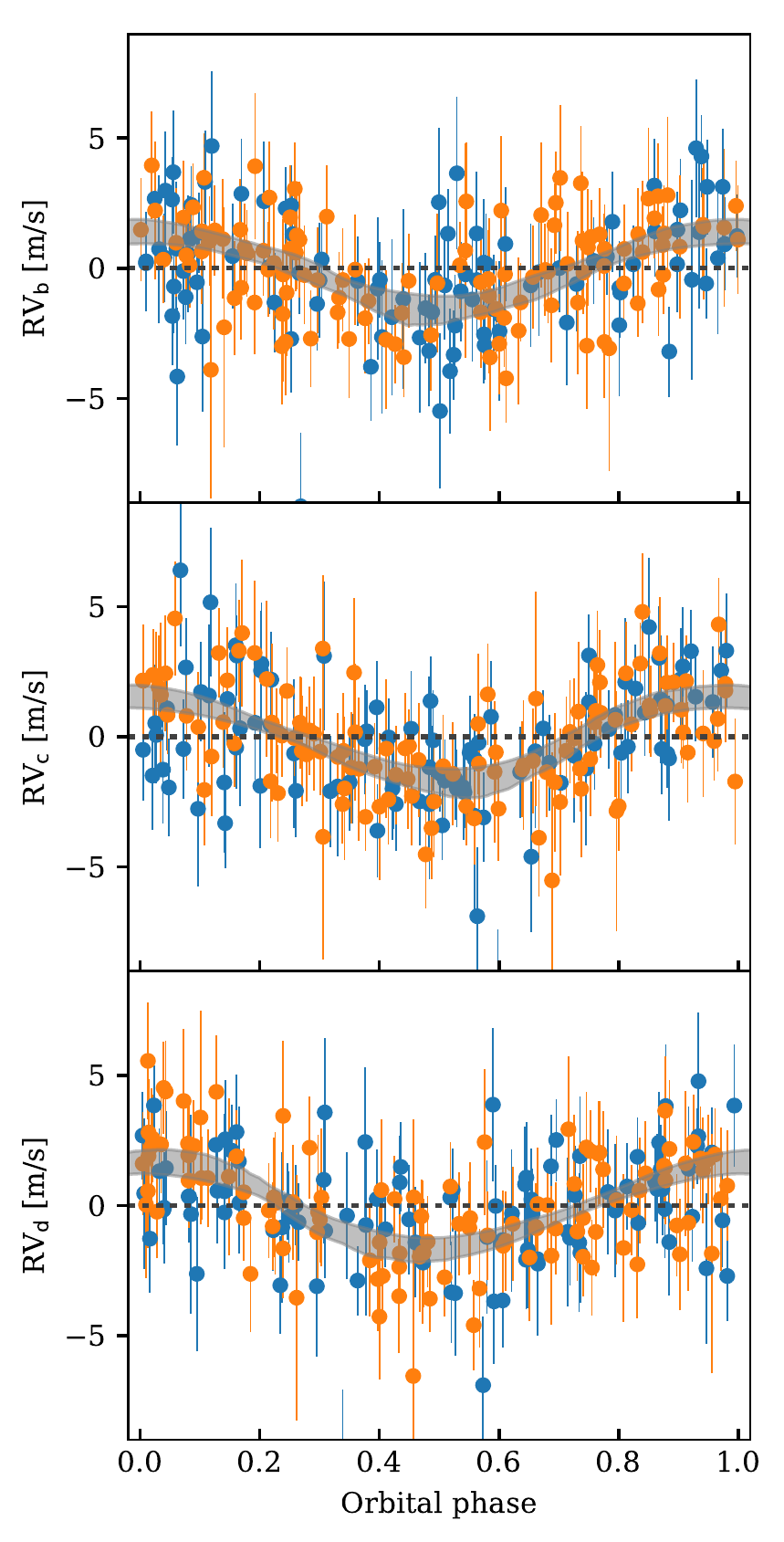}
\caption{\small Radial velocities folded to each maximum a posteriori
orbital period (planets b, c, and d from top to bottom, respectively).
Gray shaded regions extend between the 5th and 95th percentile of the
model at each orbital phase, computed over 10,000 samples of the
posterior. Orange and blue circles depict data before and after the 
HARPS fibre upgrade, respectively.}
\label{fig:phaserv}
\end{figure}

The periodogram and activity analysis (Sect.~\ref{subsec:kep},
\ref{subsec:act}) both give strong evidence that the very-low-mass
star
YZ~Cet is orbited by at least three planets in a compact
architecture. From the Keplerian modeling combined with the 
Gaussian processes, we obtain orbital periods of 4.66, 3.06, and 1.97
days, that is, orbits close to the 3:2 mean-motion resonances.
Considering the stellar mass of 0.13M$_\odot$, the measured
semi-amplitude translates into terrestrial-mass planets with minimum
masses of 1.14, 0.98, and 0.75 Earth-masses, respectively. 
Table~\ref{tab:OrbParamSmall} gives the basic parameters of the
system. Figure~\ref{fig:phaserv} shows the RV folded to each period.

\begin{table}
\centering
\caption{Basic orbital parameters of the planets. All  
parameters can be found in Table~\ref{tab:OrbitalParam}.}
\label{tab:OrbParamSmall}
\begin{tabular*}{\hsize}{@{\extracolsep{\fill}} l c c c c}

\noalign{\smallskip}
\hline
\noalign{\smallskip}
   &  &\bf YZ~Cet b&\bf YZ~Cet c & \bf YZ~Cet d\\
\noalign{\smallskip}
\hline
\noalign{\smallskip}
M $\sin$ i & [M$_{\rm Earth}$] & $0.75\pm0.13$ & $0.98\pm0.14$ & $1.14\pm0.17$\\
\noalign{\smallskip}
 P &[days]    &  $1.96876$&    $3.06008$& $4.65627$\\
  & [days] &   $\pm0.00021$&    $\pm0.00022$& $\pm0.00042$\\
\noalign{\smallskip}
e & &    $0.00\pm0.12$ & $0.04\pm0.11$ & $0.129\pm0.096$ \\
\noalign{\smallskip}
a & [au] &  $0.01557$ & $0.02090$ & $0.02764$\\
 & [au] & $\pm0.00052$ & $\pm0.00070$ & $\pm0.00093$\\

\noalign{\smallskip}
\hline

\end{tabular*}
\end{table}

\subsection{Dynamical modeling}
\label{subsec:dyn}

\begin{figure}[t]
\centering
\includegraphics[width=\linewidth]{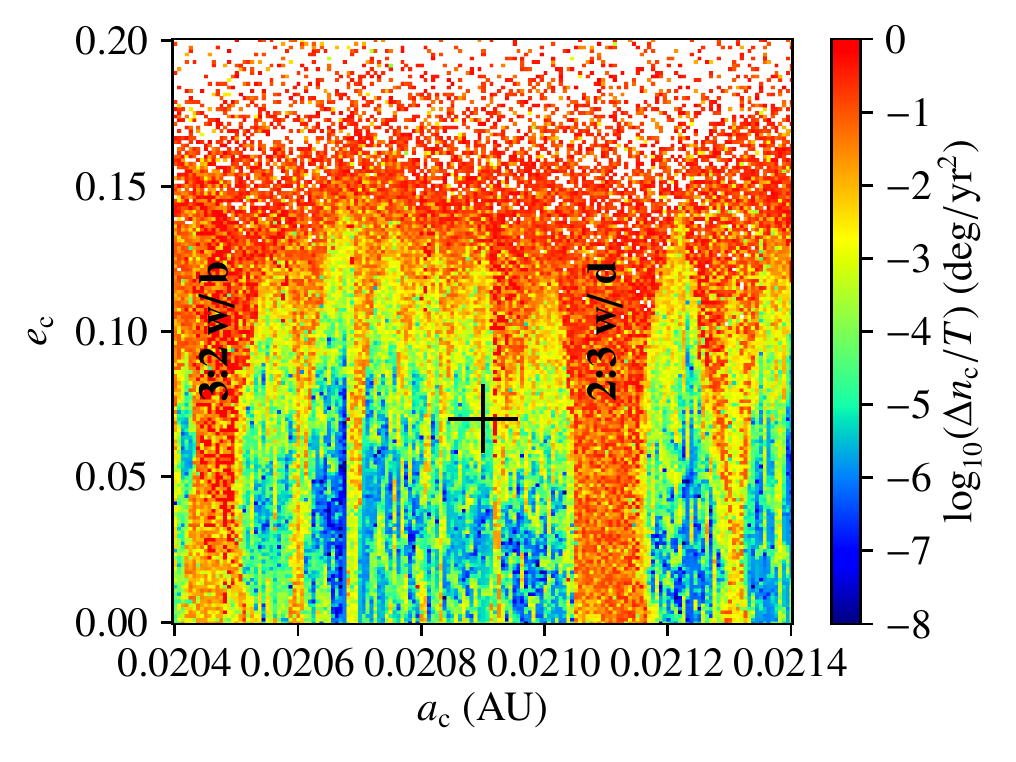}
\caption{\small Stability analysis around the MAP solution (marked
with a black cross, see Table~\ref{tab:OrbitalParam}). Blue points
correspond to stable and red points to unstable solutions. White
points correspond to highly unstable solutions, i.e., with a collision
or the ejection of a planet during the 1~kyr of the simulation. The
3:2 resonances between planets b and c and between planets c 
and d are highlighted.}
\label{fig:stab}
\end{figure}

We show in Figure~\ref{fig:stab} a stability analysis focused on
planet c where all orbital parameters are fixed to the
maximum a posteriori (MAP) solution, and we varied only the semi-major 
axis and the eccentricity. For each set of initial conditions, we 
integrated the system for 1~kyr using GENGA
\citep{grimm_genga_2014}, and computed the NAFF stability indicator
\citep{laskar_chaotic_1990,laskar_frequency_1993,correia_coralie_2005}. 
This showed that the MAP solution is stable and outside the two 3:2 
mean-motion resonances. We also see that $e_\mathrm{c} \lesssim 0.1$
is required for the system to remain stable.

Additionally, we used the n-body code \reb\ \citep{Rein2012} with the 
\whf\ integrator \citep{Rein2015} to obtain the RV of the bodies in time, which we compared with the HARPS data. A stability criterion was imposed 
based on an additional n-body integration of 1~kyr in the future.
The change in period of a planet in 1~kyr ($\Delta P_{\rm 1~kyr}$)
was extrapolated to 5~Gyr (the estimated age of the system) assuming
$\Delta P_{\rm 5~Gyr}\sim\Delta P_{\rm 1~kyr}\sqrt{\rm 5~Gyr/1~kyr}$.
The model was rejected when $\Delta P_{\rm 5~Gyr}$ was longer than the
measured orbital period, which is a conservative criterion 
rejecting unstable configurations. We used a normal distribution as
prior for the stellar mass (Table~\ref{tab:stellarprop}) and uniform
priors for the remaining parameters. The \emcee\ algorithm
\citep{goodmanweare2010, emcee} was used to sample the
posterior distribution of the parameters.
We estimated the planet masses without the degeneracy with an orbital
inclination relative to the line of sight by modeling the gravitational
interaction between the three well-known  planets in the system. From 
this preliminary analysis, until the fourth signal is confirmed
(or rejected), we obtain 
M$_{\rm b} = 1.30^{+1.0}_{-0.56}$M$_\oplus$, M$_{\rm c} =
2.1^{+1.5}_{-1.0}$M$_\oplus$, and M$_{\rm d} = 1.74^{+1.3}_{-0.51}
$M$_\oplus$ (median and 68.3\% credible interval), thanks to the
constraint on the inclination: 
95\% HDI $i_{\rm b}$[10, 172]\degree, $i_{\rm c}$[6,
174]\degree, and $i_{\rm d}$[28, 174]\degree.  Tighter constraints
are also obtained for the eccentricities of the measured planets, 
$e_{\rm b}<0.39$, $e_{\rm c}<0.15$, and $e_{\rm d}<0.17$ at 95\%.

\section{Conclusions}
\label{sec:conclusion}

We presented the discovery of three terrestrial-mass planets
orbiting the very-low-mass star YZ~Cet. The best fit (three
Keplerian+GP) to our 13 years of HARPS data results in compact orbits
close to the 3:2 mean-motion resonance with orbital periods of 1.97,
3.06, and 4.66 days. The measured minimum masses are 0.75, 0.98, and
1.14 Earth-masses, respectively, with a typical uncertainty of 16\%.
A fourth planet with a minimum mass of 0.42$\pm$0.11M$_\oplus$ and
an orbital period of 1.0419 days may be present in the system (1.1\%FAP). 
To date, these are the lowest minimum masses measured through radial
velocity. 

From a dynamical analysis, we infer the upper limits of the planetary
masses, resulting in masses below three Earth-masses. It is probable 
that these have radii up to 1.5R$_\oplus$ \citep{Fulton2017} with a
rocky bulk composition. The planets orbit outside the habitable
zone of YZ~Cet, with equilibrium temperatures ranging from 347--491 K,
299--423 K, and 260--368 K for planets b, c, and d, respectively.

The the system is at only 3.6 pc, making it very attractive for
further characterization, especially if any of the planets is caught
transiting its host star, which is not the case so far.

\begin{acknowledgements}
NAD and FP acknowledges the support of the Swiss National Science
Foundation project 166227. NCS acknowledges support from Funda\c{c}\~ao
para a Ci\^encia e a Tecnologia (FCT) through national funds and from
FEDER through COMPETE2020 by the grants UID/FIS/04434/2013 \&
POCI--01--0145-FEDER--007672 and PTDC/FIS-AST/1526/2014 \&
POCI--01--0145-FEDER--016886. NCS also acknowledges the support from FCT
through Investigador FCT contract IF/00169/2012/CP0150/CT0002.
We made use of the \reb\ code, which can be downloaded at 
\url{http://github.com/hannorein/rebound}. These simulations have been 
run on the {\it Regor} cluster kindly provided by the Observatoire de
Gen\`eve. 
\end{acknowledgements}

\bibliographystyle{aa}
\bibliography{Gl54d1}

\begin{appendix}

\section{Gaussian process model}
\label{sec:GP}

A quasi-periodic kernel was chosen to construct the elements of the
covariance matrix of the observations. This type of covariance
function was used in the past to model the effect of stellar active
regions rotating in and out of view as the star revolves
\citep{Haywood2014, Rajpaul2015}.

The elements of the covariance matrix are then given by $K_{ij} = K^r_{ij} + K^w_{ij}$, where 
\begin{equation*}
    K^r_{ij} = k_{\{A;\tau;\per;\struct\}}(t_i, t_j) = A^2 \exp\left[-\frac{(t_i - t_j)^2}{2 \tau^2} - \frac{2}{\struct} \sin^2\left(\frac{\pi (t_i - t_j)}{\per}\right)\right]
    \label{eq.kernel}
\end{equation*}
represents the RV effect related to the stellar
rotational modulation, with $\{A,\tau,\per,\struct\}$ the
hyperparameters associated with the covariance amplitude, the
evolution timescale, the recurrence timescale (the rotational period
of the star), and the smoothing parameter, respectively. The second
term is a diagonal matrix with the quadratic sum of the internal
velocity errors ($\sigma_i$) and an additional white-noise component: 
\begin{equation*}
    K^w_{ij} = \left(\sigma_i^2 + S_i\sigma_J^2 + S^+_i\sigma^{+2}_J\right) \;\delta_{ij}\;\;,
\end{equation*}
where $S_i$ is an indicator variable that is one if observation $i$
is taken before the HARPS fibre upgrade and zero otherwise, and
vice versa for $S^+_i$. $\delta_{ij}$ is the Kronecker delta function.

The prior distribution of the hyperparameters can be factorized in
the marginal prior distributions described in
Table~\ref{table.priorhyp}. Only the recurrence timescale $\per$ was
assigned an informative prior based on the analysis of the ASAS
photometric measurements and the FWHM time series. 

\section{Details on the MCMC analysis}
\label{sect:MCMC}

The ensemble sampler of \citet{goodmanweare2010} implemented in the
emcee package \citep{emcee} was used to sample the joint posterior
distribution of the parameters and hyperparameters of the model. The
priors assigned to the model parameters are listed in Tables
\ref{table.priors} and \ref{table.priorhyp}.

Initially, we chose to use 100 walkers on a 23-dimensional parameter
space. The starting position of each walker was randomly drawn from the
prior distribution. The burn-in period took approximately 500,000 steps,
and the walkers were allowed to evolve still for around 100,000
steps. An issue usually encountered when running an MCMC algorithm is
that some walkers become stuck in the local maxima, in this case, related
with period aliases and other features seen in the periodogram
(Fig.~\ref{fig:rv_period}). Because of this, we decided to retain
only those walkers that explored the main periodogram peak for each
planet. In this manner, 47 out of the 100 walkers were kept. The samples
produced by these subsets of walkers were used to initialize a second MCMC
run using 300 walkers, which were run for 70,000 steps. This run had a
mean acceptance fraction between walkers of around 25\%. The
characteristic autocorrelation length was between $3\times 10^2$ and
$1\times10^3$ steps. The average autocorrelation length over walkers was
shorter than 1,000 steps for all parameters. We have then more than
700 independent samples per walker, which leads to a total of around
21,000 independent samples that are used for the parameter
inference.

\begin{figure*}[t]
\centering
\includegraphics[width=\linewidth]{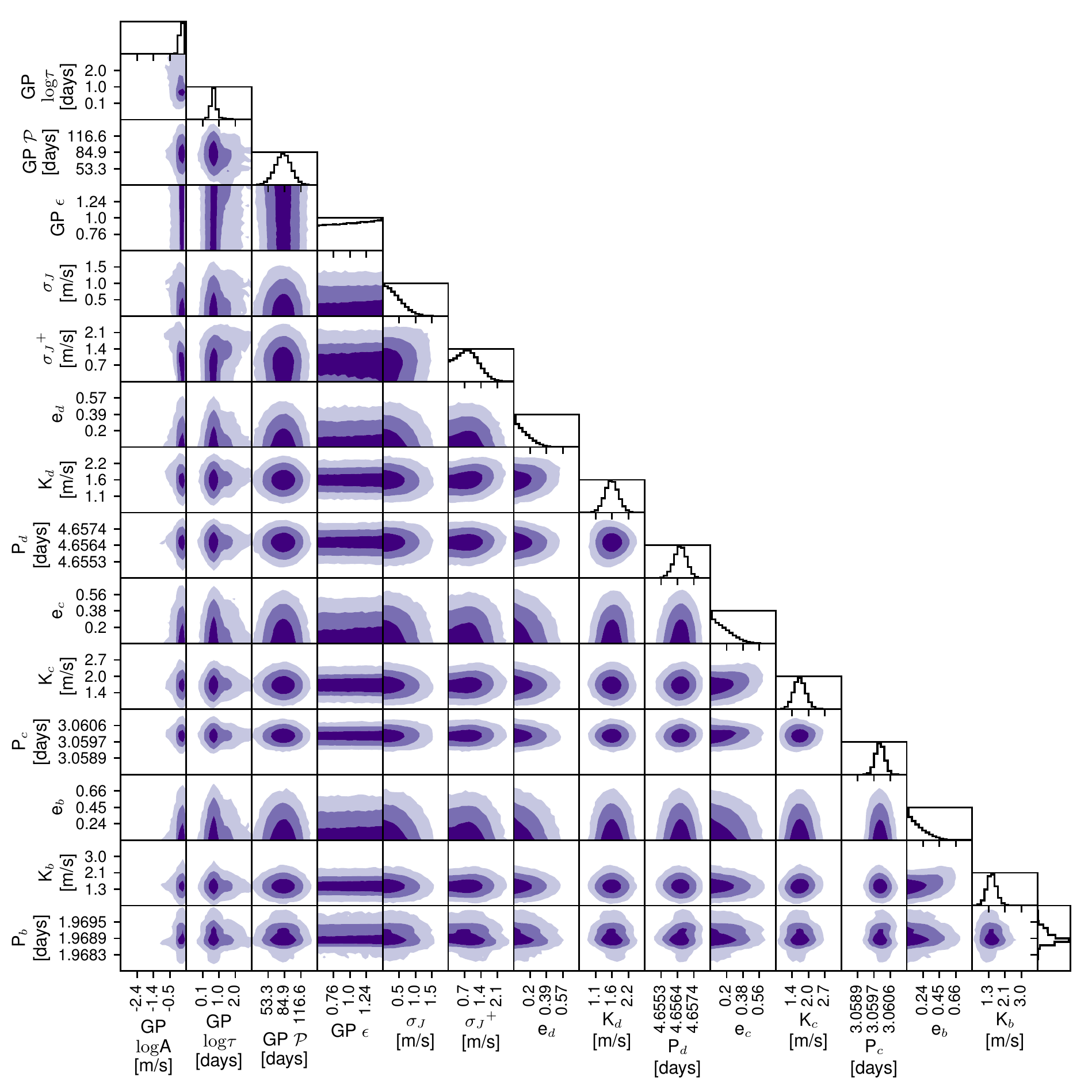}
\caption{\small One- and two-dimensional marginal posterior distributions of selected model parameters. The contours represent the 68.27\%, 95.45\%, and 99.73\% joint credible regions.}
\label{fig:pyramid}
\end{figure*}

In Table~\ref{tab:OrbitalParam} we list some characteristics of the
posterior distribution based on the MCMC samples.
Figure~\ref{fig:pyramid} shows the 2D marginal distributions of a
selection of model parameters.

\newcommand{\jeff}{\mathcal{J}}
\newcommand{\unif}{\mathcal{U}}

\begin{table}[h]
\caption{Prior distribution for the Keplerian curves and velocity
offset between HARPS pre- and post-fibre upgrade. The priors are the
same for each Keplerian, except for the period parameter.
$\unif(x_{min}, x_{max})$ is the uniform distribution between
$x_{min}$ and $x_{max}$.}
    \centering
    \begin{tabular}{c|ccc}
    \hline\hline
         Parameter &  \multicolumn{3}{c}{Prior distribution}\\
         and units &YZ Cet b & YZ Cet c & YZ Cet c\\ 
         \hline
         $P$ [days]& $\unif(1.5, 2.5)$ & $\unif(2.5, 3.5)$ & $\unif(4.0, 5.2)$\\
         $K$ [m/s]& \multicolumn{3}{c}{$\longleftarrow\unif(0, 10)\longrightarrow$}\\
         $\sqrt{e}.\sin({\omega})$& \multicolumn{3}{c}{$\longleftarrow\unif(-1, 1)\longrightarrow$}\\
         $\sqrt{e}.\cos({\omega})$& \multicolumn{3}{c}{$\longleftarrow\unif(-1, 1)\longrightarrow$}\\
         $\lambda_0$ &\multicolumn{3}{c}{$\longleftarrow\unif(0, 2\pi)\longrightarrow$}\\
         $e$ &\multicolumn{3}{c}{$\longleftarrow\unif(0, 1)\longrightarrow$}\\
         \hline
         $\Delta V_{21}$ [m/s] &\multicolumn{3}{c}{$\unif(-10, 10)$}\\
         \hline

    \end{tabular}
    \label{table.priors}
\end{table}

\begin{table}[h]
\caption{Prior distribution for the hyperparameters of the Gaussian process kernel function. $\jeff(x_{min}, x_{max})$ is the Jeffreys distribution (log-flat), $\unif(x_{min}, x_{max})$ is the uniform distribution between $x_{min}$ and $x_{max}$.}
    \centering
    \begin{tabular}{c|c}
    \hline\hline
         Parameter &  Prior distribution\\
         \hline
         A & $\jeff(10^{-4}\:\textrm{m/s}, 10\:\textrm{m/s})$\\
         $\tau$ & $\jeff(0.1\:\textrm{d}, 1000\:\textrm{d})$\\
         $\per$ &N(83 d, 15 d)\\
         $\epsilon$ & $\unif(0.5, 1.5)$ \\
         $\sigma_J$ & $\unif(0\:\textrm{m/s}, 10\:\textrm{m/s})$\\
         $\sigma^+_J$ & $\unif(0\:\textrm{m/s}, 10\:\textrm{m/s})$\\
         \hline
    \end{tabular}
    \label{table.priorhyp}
\end{table}

\begin{table*}[t]
\small
\caption{\small Orbital elements using a model of three Keplerian plus a 
Gaussian process. Values correspond to the maximum-a-posteriori 
of the MCMC, with errors being the standard deviation of the MCMC samples. In the second line we report the 95\% highest-density interval.}
\begin{tabular*}{\hsize}{@{\extracolsep{\fill}}llccc}
\hline\noalign{\smallskip}
N$_{\rm Meas}$&\multicolumn{4}{c}{211}\\
\noalign{\smallskip}\hline\noalign{\smallskip}
$\gamma$\qquad [m/s] & \multicolumn{1}{c}{$28301.00\pm 0.30$} &\multicolumn{1}{c|}{}& $\Delta V_{21}$\qquad [m/s]& $0.01 \pm 0.60$\\
               & \multicolumn{1}{c}{$[28300.37, 2831.62]$} & \multicolumn{1}{c|}{}&& \multicolumn{1}{c}{$[-1.62, 0.85]$}\\
\noalign{\smallskip}\hline\noalign{\smallskip}
&\multicolumn{4}{c}{GP hyperparameters}\\
\noalign{\smallskip}\hline\noalign{\smallskip}
$\sigma_J$ \hspace{35pt} [m/s] & \multicolumn{1}{c}{$0.13\pm 0.28$} &\multicolumn{1}{c|}{}& \multicolumn{1}{l}{$\sigma^+_J$ \hspace{35pt} [m/s]} & $0.64\pm 0.46$\\
                & \multicolumn{1}{c}{$[0.00, 0.95]$} &\multicolumn{1}{c|}{}&& $[0.00, 1.64]$\\

$\log_{10}(A)$ \hspace{15pt} [m/s] & \multicolumn{1}{c}{$0.14\pm 0.12$} &\multicolumn{1}{c|}{}& \multicolumn{1}{l}{$\per$ \hspace{40pt} [d]} & $79\pm17$\\
                & \multicolumn{1}{c}{$[-0.021, 0.348]$} &\multicolumn{1}{c|}{}&&$[51, 114]$\\

$\log_{10}(\tau)$ \hspace{18pt} [d] & \multicolumn{1}{c}{$0.68\pm0.27$} &\multicolumn{1}{c|}{} & \multicolumn{1}{l}{$\struct$} & $1.02\pm0.29$\\
                & \multicolumn{1}{c}{$[0.28, 1.40]$} &\multicolumn{1}{c|}{}&&$[0.54, 1.50]$\\

\noalign{\smallskip}\hline\noalign{\smallskip}
    &          &\bf YZ~Cet b&\bf YZ~Cet c & \bf YZ~Cet d\\
\noalign{\smallskip}\hline\noalign{\smallskip}
  $P$& [days]&    $1.96876\pm0.00021$&    $3.06008\pm0.00022$& $4.65627\pm0.00042$\\
                &&$[1.96858, 1.96934]$& $[3.05958, 3.06051]$&$[4.65565, 4.65740]$\\
\noalign{\smallskip}

$K$& [m/s]&    $1.48\pm0.25$ & $1.68\pm0.23$ & $1.72\pm0.23$\\
            && $[0.92, 1.92]$ & $[1.22, 2.13]$ & $[1.13, 2.09]$\\
\noalign{\smallskip}

  $e$& [0..1[&    $0.00\pm0.12$ & $0.04\pm0.11$ & $0.129\pm0.096$ \\
            && $[0.00, 0.41]$ & $[0.00, 0.37]$ & $[0.000, 0.311]$\\
\noalign{\smallskip}

$\lambda_{0}$ at BJD$_{\rm ref}$=2456845.73& [deg]&    $233.9\pm9.4$ & $298.8\pm8.3$ & $319.9\pm8.8$\\
            &&  $[215.0, 254.2]$    & $[283.5, 316.4]$  & $[297.8, 333.4]$\\
\noalign{\smallskip}\hline\noalign{\smallskip}

$\sqrt{e}.\cos({\omega})$&  &   $0.02\pm0.27$ &  $-0.19\pm0.21$ & $-0.06\pm0.22$\\
                            &&  $[-0.61, 0.43]$ & $[-0.48, 0.32]$ & $[-0.41, 0.43]$ \\  
\noalign{\smallskip}

$\sqrt{e}.\sin({\omega})$&  &   $-0.04\pm0.26$ & $-0.07\pm0.24$ & $0.36\pm0.22$\\
                            &&  $[-0.47, 0.56]$ & $[-0.63, 0.26]$ &$[-0.28, 0.55]$\\
\noalign{\smallskip}

M $\sin$ i& [M$_{\rm Earth}$]&   $0.75\pm0.13$ & $0.98\pm0.14$ & $1.14\pm0.17$\\
                           && $[0.436, 0.964]$ & $[0.675, 1.267]$ & $[0.723, 1.389]$\\
\noalign{\smallskip}

  a& [au]&                  $0.01557\pm0.00052$ & $0.02090\pm0.00070$ & $0.02764\pm0.00093$\\
        &&      $[0.01444, 0.01661]$ & $[0.01938, 0.02229]$ & $[0.02564, 0.02948]$\\

\noalign{\smallskip}
  T$_{eq}$ for A$_B$=[0.75, 0]& [K] & [347, 491] & [299, 423] & [260, 368]\\
\noalign{\smallskip}

  BJD$_{\rm Trans}-2456846.0$&[days]  & $0.911\pm0.095$ & $1.05\pm0.14$ & $1.44\pm0.16$ \\
        &&      $[0.723, 1.137]$ & $[0.74, 1.35]$ & $[1.16, 1.79]$\\
  
\noalign{\smallskip}
  Trans. Prob. & [\%] & 5.7 & 4.1 & 2.9\\
\noalign{\smallskip}
\hline
\end{tabular*}
\label{tab:OrbitalParam}
\end{table*}

\begin{table*}
\caption{Radial velocity time series for YZ~Cet (minimal; full version available at the CDS).}
\label{tab:rv}
\begin{tabular*}{\hsize}{@{\extracolsep{\fill}}lllllllll}

\noalign{\smallskip}
\hline\hline
\noalign{\smallskip}

BJD - 2400000 & RV &  $\sigma_{RV}$ & FWHM & Contrast & BIS & S-index & $H\alpha$\\
 & $[kms^{-1}]$ & $[kms^{-1}]$ & [$kms^{-1}$] & & [$kms^{-1}$] &&  \\

\noalign{\smallskip}
\hline
\noalign{\smallskip}

52986.601206 & 28.30238 & 0.00411 & 3.88737 & 32.84564 & -6.56400  & 52.61462 & 0.10798\\
52996.557553 & 28.29836 & 0.00191 & 3.85895 & 32.12375 & -16.80400 & 5.11522  & 0.13170\\
53337.624714 & 28.29440 & 0.00126 & 3.83507 & 32.66167 & -20.33000 & 4.52328  & 0.12505\\
53572.915467 & 28.30120 & 0.00196 & 3.84791 & 32.46511 & -18.21900 & 5.86166  & 0.12981\\
53573.902402 & 28.29976 & 0.00271 & 3.85660 & 32.13906 & -24.79800 & 5.77385  & 0.12642\\

\noalign{\smallskip}
\hline

\end{tabular*}
\end{table*}

\end{appendix}

\end{document}